\pdfoutput=1

\RequirePackage{ifpdf}
\documentclass[12pt,nohyper]{JHEP3}

\usepackage{epsfig}
\usepackage{float}
\usepackage{amsmath}
\usepackage{array}
\usepackage{cite}
\usepackage{arydshln}

\newcommand{\smallminus}{{\rm\rule[2.4pt]{6pt}{0.65pt}}}
\newcommand{\smallplus}{\hspace{0.5pt}\text{{\small+}}\hspace{-0.5pt}}
\newcommand{\mi}{\smallminus}
\newcommand{\pl}{\smallplus}
\newcommand{\la}{\langle}
\newcommand{\ra}{\rangle}

\title{\hspace{-0.0cm}{\LARGE Positive Amplitudes In The Amplituhedron}}
\author{\vspace{-.5cm}Nima Arkani-Hamed$^{a}$, Andrew Hodges$^{b}$ and Jaroslav Trnka$^{c}$\\

{\footnotesize{\it $^{a}$ School of Natural Sciences, Institute for Advanced Study, Princeton, NJ 08540, USA}\\
{\it $^{b}$ Wadham College, University of Oxford, Oxford OX1 3PN, UK}\\
{\it $^{c}$ Walter Burke Institute for Theoretical Physics, California Institute of Technology, Pasadena, CA 91125,
USA}} }

\preprint{2014}

\abstract{The all-loop integrand for scattering amplitudes in planar ${\cal N}=4$ SYM is determined by an ``amplitude form" with logarithmic singularities on the boundary of the amplituhedron. In this note we provide strong evidence for a new striking property of the superamplitude, which we conjecture to be true to all loop orders: the amplitude form is positive when evaluated inside the amplituhedron. The statement is sensibly formulated thanks to the natural ``bosonization" of the superamplitude associated with the amplituhedron geometry.  However this positivity is not manifest in any of the current approaches to scattering amplitudes, and in particular not in the cellulations of the amplituhedron related to on-shell diagrams and the positive grassmannian. The surprising positivity of the form suggests the existence of a ``dual amplituhedron" formulation where this feature would be made obvious. We also suggest that the positivity is associated with an extended picture of amplituhedron geometry, with the amplituhedron sitting inside a co-dimension one surface separating ``legal" and ``illegal" local singularities of the amplitude. We illustrate this in several simple examples, obtaining new expressions for amplitudes not associated with any triangulations, but following in a more invariant manner from a global view of the positive geometry.}

\preprint{CALT-TH-2014-168}

\begin{document}

\section{Introduction}

The amplituhedron ${\cal A}_{n,k,L;m}$ \cite{Arkani-Hamed:2013jha} (see also \cite{Arkani-Hamed:2013kca,Franco:2014csa,Bai:2014cna,Lam:2014jda} for recent developments) lives in $G(k,k+m;L)$, which is the space
of $k$-planes $Y$ in $k+m$ dimensions, together with $L$ 2-planes
${\cal L}_1, \cdots, {\cal L}_L$  in the $m$-dimensional complement
of $Y$.
$$
\includegraphics[scale=0.65]{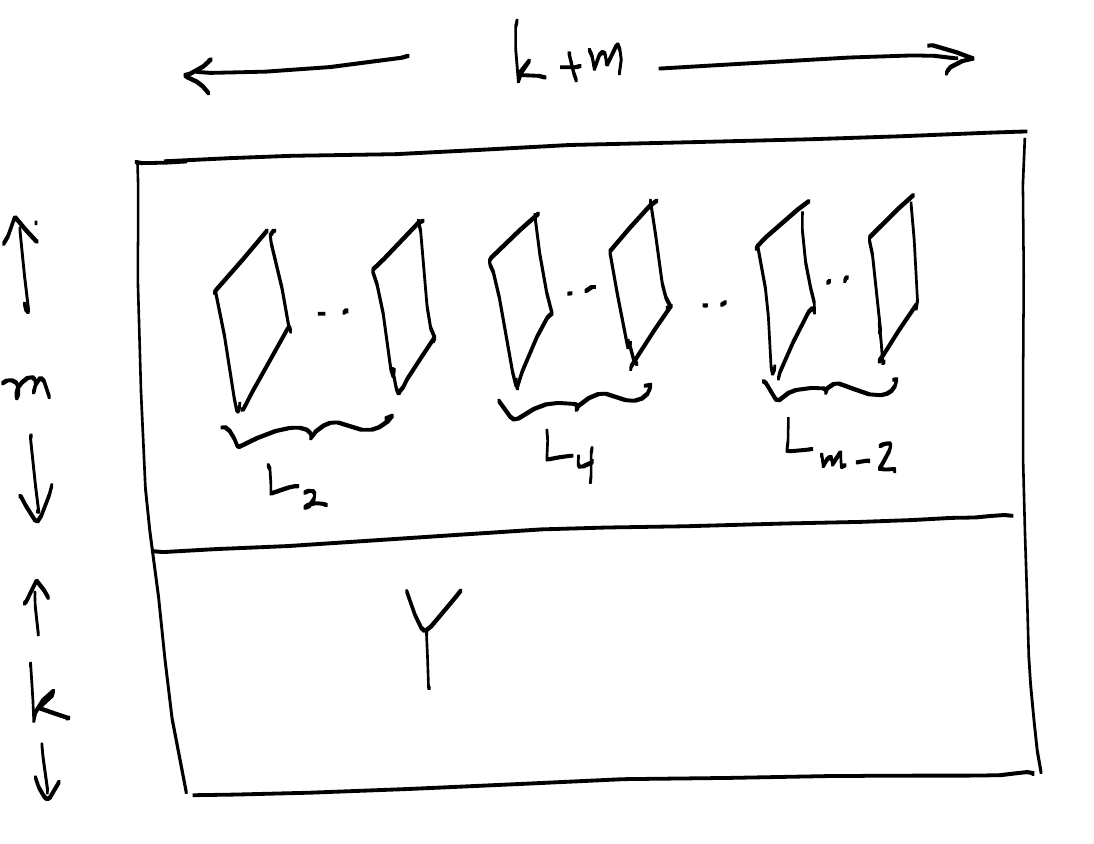}
$$
The ``external data" are given by $n$ $(k+m)$-dimensional vectors $Z_a^I$, where $a=1, \cdots n$, and $I = 1,
\cdots, (k+m)$. This data is ``positive": the ordered $(k+m) \times (k+m)$ determinants $\langle
Z_{a_1} \cdots Z_{a_{k+m}} \rangle > 0$ for $a_1 < \cdots <
a_{k+m}$. The subspace of ${\cal A}_{n,k,L;m}$ of $G(k,k+m;L)$ is
determined by a ``positive" linear combination of this positive
external data.  The $k$-plane is $Y_{\alpha}^I$, and the 2-planes
are ${\cal L}_{\gamma (i)}^I$, where $\gamma = 1,2$ and  $i = 1,
\dots, L$. We will often refer to these in combination as ${\cal Y}$. The amplituhedron is the space of all ${\cal Y}$ of the form
\begin{equation}
{\cal Y} = {\cal C} \cdot Z
\end{equation}
or more explicitly
\begin{equation}
Y_{\alpha}^I = C_{\alpha a} Z_a^I, \qquad {\cal L}_{\gamma (i)}^{I}
= D_{\gamma a (i)} Z_a^I
\end{equation}
Here $C_{\alpha a}$ specifies a $k$-plane in $n$ dimensions,
and the $D_{\gamma a (i)}$ are $L$ 2-planes living in the $(n-k)$-dimensional complement of $C$. 
$$
\includegraphics[scale=0.65]{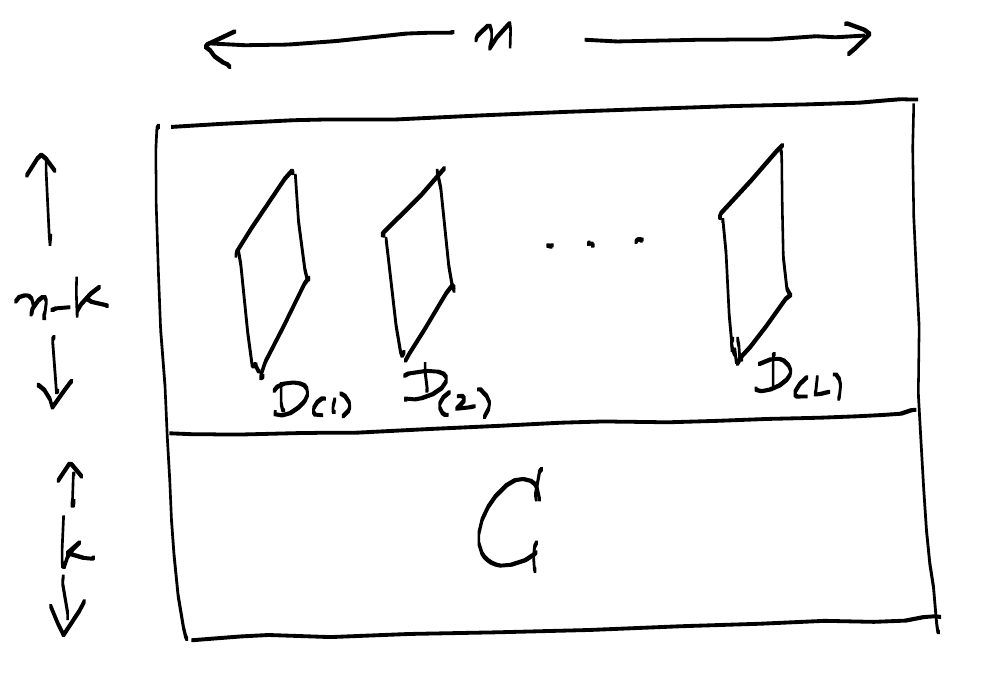}
$$
The $C,D$ matrices have the positivity property that for
any $0 \leq l \leq L$, all the ordered maximal minors of the $(k + 2
l) \times n$ matrix
\begin{equation}
\left(\begin{array}{ccc} & D_{(i_1)} & \\
\hdashline & \vdots &  \\ \hdashline & D_{(i_l)} & \\ \hdashline & C
\end{array} \right)
\end{equation}
are positive.

The existence of the amplituhedron was strongly motivated by the ``polytope picture" of \cite{Hodges:2009hk}; the amplituhedron explains the origin of these polytopes and extends the story for all $k$ and to all loop orders. The amplituhedron most directly relevant for scattering amplitudes in planar ${\cal N}=4$ SYM has $m=4$. The superamplitude is extracted from a canonical form $\Omega_{n,k,L}[{\cal Y},Z]$, with logarithmic singularities on the boundary of the amplituhedron. One approach to determining $\Omega$ begins with ``triangulating" or ``cellulating" the space \cite{ArkaniHamed:2012nw,Arkani-Hamed:2013jha}. However this is not a completely satisfactory approach, and we would prefer to have a more invariant definition of $\Omega$.

We do have a more satisfactory picture for determining $\Omega$ at least for $k=1$ and any $m$, where the amplituhedron is a cyclic polytope in $\mathbb{P}^{k+m-1}$. $\Omega$ can be described either as the form with logarithmic singularities on the boundary of ${\cal A}$, or writing $\Omega = \langle Y d^m Y \rangle f(Y)$, we can think of $f(Y)$ as the literal volume of the {\it dual} of the amplituhedron $\widetilde{{\cal A}}$. In this case, we can write $\Omega$ as an integral over the space of planes $W$ dual to the points $Y$ as
\begin{equation}
\Omega = \langle Y d^m Y \rangle \int_{\tilde{{\cal A}}} \frac{\langle W d^m W \rangle}{(W \cdot Y)^{m+1}}
\end{equation}
as given in \cite{Hodges:2009hk} and \cite{ArkaniHamed:2010gg}.

While do not yet know what the ``dual amplituhedron" might mean for $k>1$, or what the analog of the above integral representation might look like, we suspect that finding this dual formulation will be the missing ingredient needed to make contact between the beautiful geometric structures seen in the integrand and the emergence of a worldsheet description turning into the weakly coupled string at strong 't Hooft coupling.

Our purpose in this note is instead to give strong evidence that some second formulation of this type exists, by observing a remarkable new feature of the form $\Omega$ which we conjecture is true for all $n,k$ at all loop orders: $\Omega$ is everywhere ${\it positive}$ when evaluated {\it inside} the amplituhedron. This is an extremely simple and striking qualitative fact about planar ${\cal N} = 4$ SYM super-integrands. Of course for general $k$, this statement is only sensible using the bosonic $Y$ space of the amplituhedron. This fact is not at all manifest in the direct triangulations of the amplituhedron, e.g. based on the BCFW expansion \cite{Britto:2004ap,Britto:2005fq,ArkaniHamed:2010kv}: individual BCFW terms can have either sign, but the sum is always positive. This is also true for $k=1$, but here, the representation of $\Omega$ as the volume of the ``dual polytope" makes the positivity manifest. 

This surprising positivity of $\Omega$ is associated with an extended understanding of the geometry of the amplituhedron. A simple feature of the amplituhedron geometry is that, say for trees, the co-dimension one boundaries occur when $\langle Y Z_i Z_{i+1} Z_j Z_{j+1} \rangle \to 0$; this also tells us that the only poles of the superamplitude are the usual local ones. Now, these co-dimension one boundaries have an extremely intricate pattern of mutual self-intersection on lower-dimensional spaces. Of course the amplitude only has non-vanishing residues on a tiny subset of these intersections; there are many more ``bad" intersections, not occurring as residues of $\Omega$, than ``good" ones. Indeed, the geometry of the amplituhedron itself tells us where the ``good" intersections are --- these are precisely those that form the boundaries of the amplituhedron. We will see below, in a number of explicit examples, something more than this: in a precise sense all the ``bad" intersections  are ``outside" the amplituhedron. This is reflected in the form $\Omega$ in an interesting way. If we write $\Omega$ as a numerator ${\cal N}$ multiplying all the local poles capturing the possible co-dimension one boundaries, we find that the ``good" singularities are separated from the ``bad" ones by a co-dimension one surface where ${\cal N}=0$.
This {\it zero surface} lies outside the positive region and only touches it on at most codimension-two boundaries. The form of this zero surface guarantees the positivity of $\Omega$ inside the amplituhedron. It follows that the form for the amplitude must be positive when it is evaluated inside the amplituhedron.

We construct this zero surface for few simple cases in section 2. This provides us with a novel picture, and hence new formulas for the amplitudes, which does not involve any sort of triangulation or representation of the amplitude as a sum of pieces, but is much more invariant, directly determining the amplitude from the global geometry of the amplituhedron.  As we will see, the geometry is quite intricate even in the simplest cases.  We have not attempted to extend this picture to general $k,L,$ though we expect it is possible to do so. Instead, in section 3 we provide evidence for the positivity conjecture  by evaluating the form $\Omega$ inside the amplituhedron and checking numerically that it is positive. In addition, we show that the positivity surprisingly seems to also hold for other objects --- the logarithm of the MHV amplitude and the ratio function. The ratio function is an IR-finite quantity and we show in a simple case that the positivity holds even after an integration has been performed to obtain the final amplitudes.

\section{Numerator As Zero Surface}
We begin by discussing the simplest classes of tree-level amplituhedra and construct their forms explicitly from a study of the allowed singularities as determined by the boundaries of the amplituhedron, starting with $m=2$ kinematics. The external data are given by $Z_1,Z_2,\dots,Z_n$, and $Y$ is a $k$-plane in $k+2$ dimensions. Amplituhedron positivity easily implies that $\langle Y Z_i Z_{i+1} \rangle >0$ inside the amplituhedron, and the co-dimension one boundaries occur when $\langle Y i \, i+1 \rangle \to 0$. Thus, a factor $\langle Y i \, i+1 \rangle$ must appear for all $i$ in the denominator of the form $\Omega_{n,k}$, and so $\Omega_{n,k}$ takes the form
\begin{equation}
\Omega_{n,k} = \frac{d\mu\,\,{\cal N}(Y)}{\la Y12\ra\la Y23\ra\la Y34\ra\dots \la Yn1\ra}
\end{equation}
where $d\mu$ denotes the standard measure $d\mu = \prod_{j=1}^k \la Yd^2Y_j\ra$, with the $k$-plane $Y$ spanned by the $k$ independent vectors $Y_1,\cdots,Y_k$.  

While the first boundaries of the space are explicitly represented in $\Omega$ by the poles arising each of the factors in the denominator, the lower-dimensional boundaries are seen by taking further residues of $\Omega$.  However only a small subset of residues given by setting $\la Y\dots\ra=\la Y\dots\ra = \dots =0$ will correspond to boundaries of the amplituhedron; most are spurious and the numerator must vanish when $Y$ approaches them. As we show in next two subsections, vanishing on all spurious boundaries is enough to determine the numerator uniquely. In these cases the explicit construction shows that all these bad boundaries are outside the amplituhedron and therefore the form $\Omega$ is positive when evaluated inside the amplituhedron. We can consider the space where ${\cal N}(Y,Z_i)=0$ as a surface of spurious points which lie outside the amplituhedron. It turns out that for $k=1$ this zero surface is specified by spurious points only while for $k=2$ it must also include spurious lines. For general $k$ it has to include projective $(k-1)$-planes.

In the last two subsections we repeat the exercise for $k=1$ with $m=3$ and $m=4$. There we find new features as the zero surface touches the positive space at points for $m=3$ and also at lines for $m=4$.

\subsection{Polygons}

The simplest case is $k=1$, where the amplituhedron is just the set of points $Y$ in $\mathbb{P}^2$ that are inside a convex polygon determined by the external data. This case was studied in detail in \cite{ArkaniHamed:2010gg}.
The first boundaries are obviously the lines $Z_iZ_{i\pl1}$ and the second boundaries are points $Z_i$. Now the denominator generates a singularity whenever we set $\la Y\,i\,i\pl1\ra=\la Y\,j\,j\pl1\ra=0$ by localizing $Y=X_{ij}$ where
\begin{equation}
X_{ij} = (i\,i\pl1)\cap(j\,j\pl1)\qquad\mbox{for $|i-j|>1,$}
\end{equation}
where $(ab)\cap(cd)\equiv Z_a \la bcd\ra - Z_b \la acd\ra$. There are exactly $\frac{n(n-3)}{2}$ of these points and the numerator is required to vanish whenever $Y=X_{ij}$ in order to cancel the pole in the denominator.
The numerator ${\cal N}(Y,Z_i)$ is a degree $(n-3)$ polynomial in $Y$,
\begin{equation}
{\cal N} (Y) =  C_{I_1I_2\dots
I_{n \mi 3}}Y^{I_1}Y^{I_2}\dots Y^{I_{ n\mi 3}} \equiv (C\cdot YY\dots Y)\, ,
\end{equation}
where $C$ is a symmetric tensor with $\frac{(n-1)(n-2)}{2} = \frac{n(n-3)}{2} + 1$ degrees of freedom. Therefore demanding that ${\cal N}(Y=X_{ij})=0$ for all $X_{ij}$ specifies the numerator completely up to an overall constant.

Let us give few examples. For $n=3$ the form is trivial as there is no $Y$ dependence in the numerator. For $n=4$ the numerator is linear in $Y$. At the same time there are two spurious points $X_{13}=(12)\cap(34)$ and $X_{24}=(23)\cap(41)$ on which the denominator of the form $\Omega$ generates a singularity.

$$
\begin{minipage}[c]{0.45\textwidth}
\includegraphics[scale=0.6]{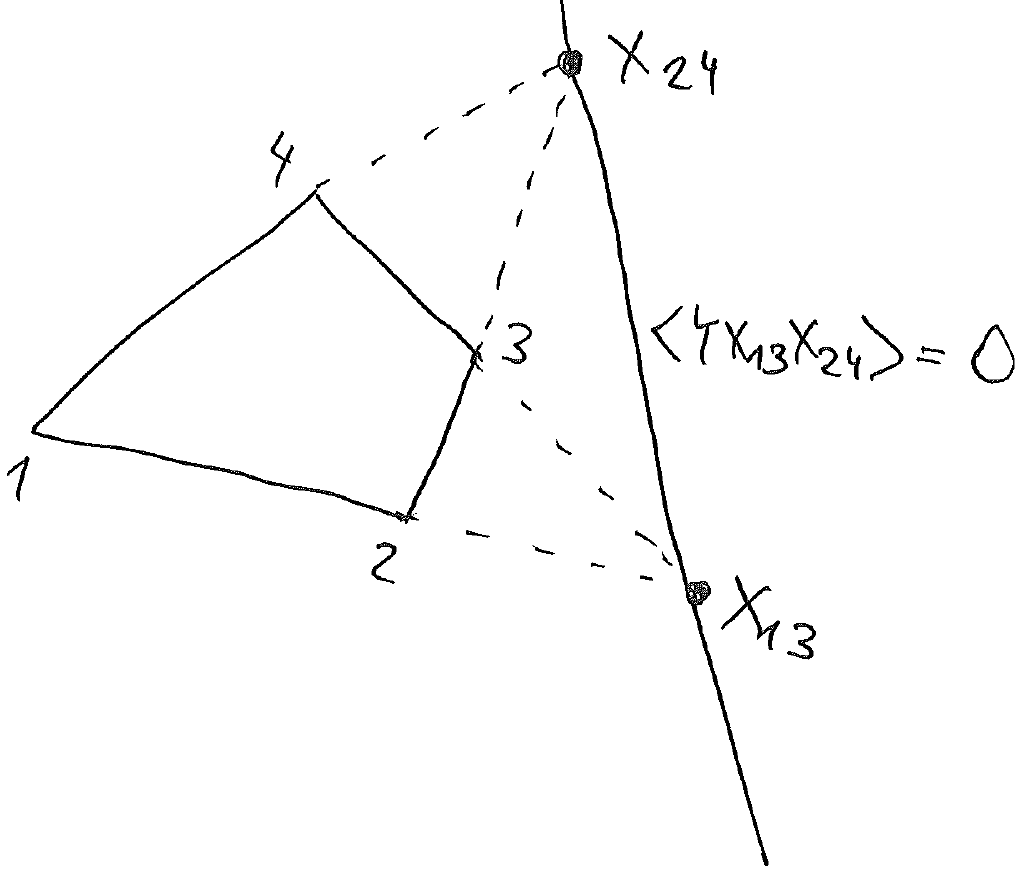}
\end{minipage}
\begin{minipage}[c]{0.55\textwidth}

It is easy to see that the form $\Omega$ is
$$
\Omega_4 = \frac{\la Y\,dY\,dY\ra\la YX_{13}X_{24}\ra}{\la Y12\ra\la
Y23\ra\la Y34\ra\la Y41\ra}
$$
where the numerator is fixed by the requirement that it vanishes when $Y$ is on the line $X_{13}X_{24}$ . Importantly this line is outside the polygon and therefore the form $\Omega_4$ is positive when evaluated inside the positive region.

\end{minipage}
$$
The polygon can be triangulated as a sum of two triangles which can be done algebraically be rewriting
$\la Y X_{13}X_{24}\ra =
\la Y 23\ra\la 341\ra\la 412\ra-\la Y41\ra\la 123\ra 234\ra$.

The next case $n=5$ is more interesting. We have five spurious points $X_{13}$, $X_{14}$, $X_{24}$, $X_{25}$ and $X_{35}$  for which the numerator, which is now quadratic in $Y$, must vanish:
\begin{equation}
{\cal N}(Y=X) = C_{IJ}\,X^IX^J = 0,\qquad \mbox{for $X = X_{13}, X_{14}, X_{24},
X_{25}, X_{35}$}\, .
\end{equation}
This is an equation for a conic defined by those five values of $X$.
$$
\includegraphics[scale=0.6]{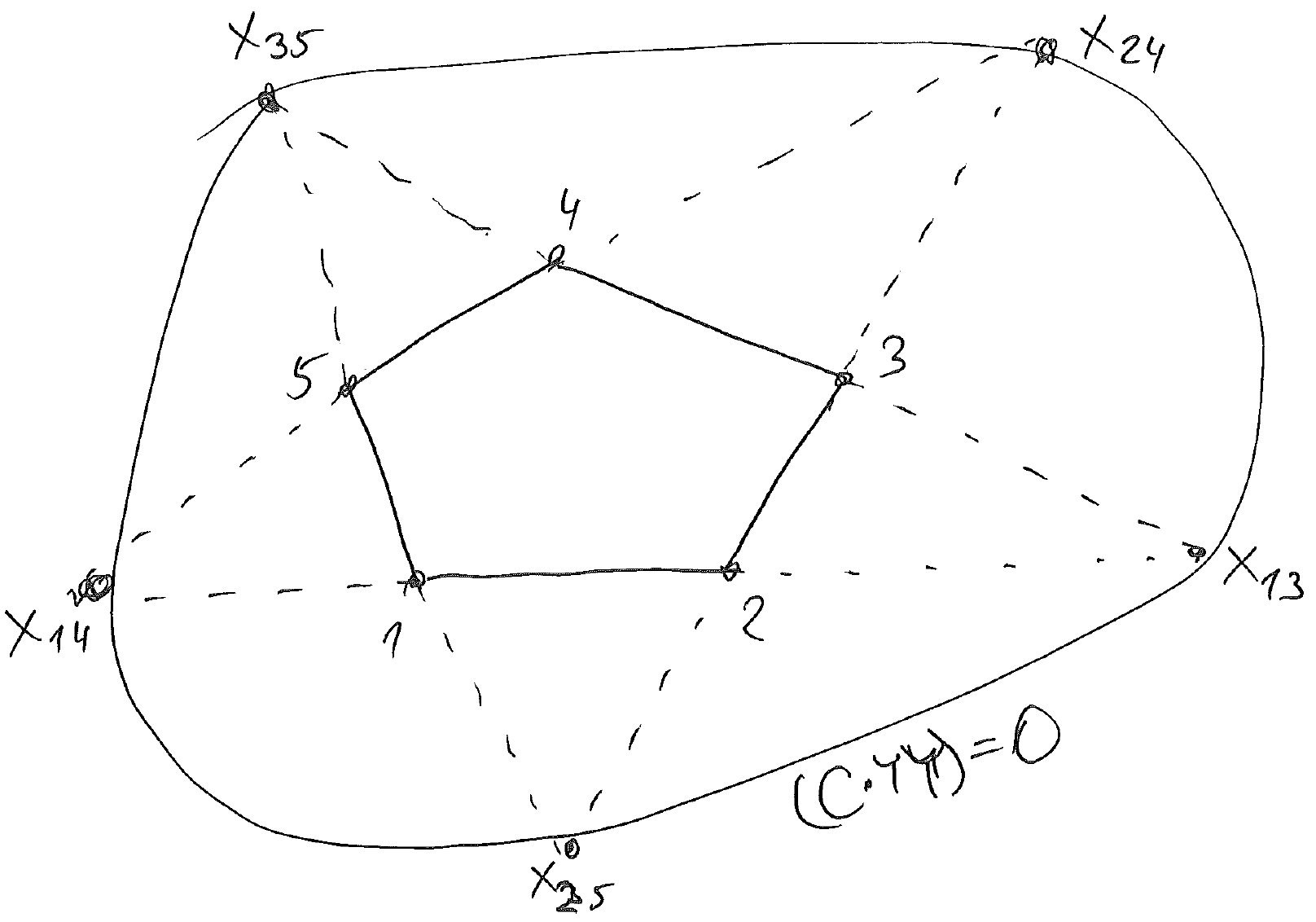}
$$
The numerator ${\cal N}(Y)$ vanishes if the point $Y$ lies on the same conic. This fixes ${\cal N}= A\cdot {\cal S}_6$ where ${\cal S}_6$ can written using the generalized $\epsilon$ symbol,
\begin{equation}
{\cal S}_6 =
\epsilon_{I_1J_1,I_2J_2,I_3J_3,I_4J_4,I_5J_5,I_6J_6}Y^{I_1}Y^{J_1}X_{13}^{I_2}X_{13}^{J_2}X_{14}^{I_3}X_{14}^{J_3}
X_{24}^{I_4}X_{24}^{J_4}X_{25}^{I_5}X_{25}^{J_5}X_{35}^{I_6}X_{35}^{J_6}
\end{equation}
where the $\epsilon$ is symmetric in $I_k$, $J_k$ and antisymmetric
in exchanging pairs $IJ$. ${\cal S}_6$ vanishes if $Y$, $X_{13}$,
$X_{14}$, $X_{24}$, $X_{25}$, $X_{35}$ lie on the same conic.
The overall constant $A$ can be then fixed by considering one leading
singularity, e.g.\ $Y=Z_1$, and demanding that  its residue has coefficient 1. This implies
\begin{equation}
A = \frac{1}{\la 123\ra\la 124\ra\la 125\ra\la 134\ra\la 135\ra\la
145\ra\la 234\ra\la 235\ra\la 245\ra\la 345\ra}\, .
\end{equation}

This is the most compact and invariant possible representation of the numerator, making all of its properties completely manifest: cyclicity and vanishing for
$Y=X_{ij}$.

The case of the general $n$-gon works in a completely analogous way. We can construct $\frac{n(n-3)}{2}$ points $X_{ij}$ which all lie outside the polygon. The numerator ${\cal N}=A_n\cdot {\cal S}_{n+1}$ is then specified by algebraic curve of degree $n-3$  which passes through all $X_{ij}$ and $Y$,
\begin{equation}
{\cal S}_{n\pl1} = \epsilon_{Y,X_{13},X_{14},\dots,X_{n\mi2\,n}}
\underbrace{Y\dots Y}_{n\mi3}\,\underbrace{\underbrace{X_{13}\dots
X_{13}}_{n\mi3}\dots\dots \underbrace{X_{n\mi2\,n}\dots
X_{n\mi2\,n}}_{n\mi3}}_{\frac{n(n-3)}{2}}
\end{equation}
where we use a collective index in the epsilon symbol to indicate a symmetric
product of the same vector. The constant  $A_n$ can be then fixed by demanding that the residues on second boundaries, i.e. on $Y=Z_i$, are $1$ (just one such check is enough).

Up to now, we have had two pictures for the form associated with the polygon. The first ``BCFW expansion" triangulates the polygon itself. We can also recognize the form as an integral expression for the area of the dual polygon, and we can find an explicit expression by triangulating the dual polygon. These two expressions make different properties of the form manifest. The BCFW triangulation of the polygon introduces interior boundaries and thus spurious poles, but only uses vertices of the polygon, so that leading singularities are at the correct locations $Z_i$ term-by-term; the positivity of the result is not manifest in each term but only arises in the sum.  The triangulation of the dual polygon has the correct poles term-by-term, but spurious locations for leading singularities that must cancel in the sum; it also makes the positivity of the form manifest. We have now given a third representation for the form, an explicit expression which does not involve breaking the polygon into triangles, and which makes {\it all} its properties obvious: the singularities are where they have to be, and the positivity is also manifest. It is amusing to find a new expression for something as elementary and familiar as the area of a convex polygon in this way, following from a more global view of the geometry, where we focus not just on the polygon itself, but also on all the ``bad" points of intersection 
$X_{ij}$ lying outside it.

\subsection{MHV 1-loop amplitude}

We move onto the case with $m=2$ and $k=2$, which is co-incidentally exactly the same geometry as $m=4$, $k=0$ and $L=1$, i.e.\ for the MHV 1-loop amplitude. Here $Y^{\alpha\beta}$ can be thought of as a line in $\mathbb{P}^3$ and the space is four-dimensional. The numerator of the form can be then written as
\begin{equation}
{\cal N}(Y) = C_{\alpha_1\beta_1,\alpha_2\beta_2,\dots,\alpha_{n-4}\beta_{n-4}}Y^{\alpha_1\beta_1}Y^{\alpha_2\beta_2}\dots Y^{\alpha_{n-3}\beta_{n-3}} \equiv (C\cdot YY\dots Y)\, .
\end{equation}
The number of degrees of freedom in $C$ is
\begin{equation}
d=2\left(\begin{array}{c}n\\4\end{array}\right)-\left(\begin{array}{c}n
\\3\end{array}\right)\label{NumD}\, .
\end{equation}
The numerator ${\cal N}(Y)$ again vanishes on a three-dimensional ``zero surface" outside the positive space.
As in previous cases the denominator of the form generates spurious singularities and therefore this zero surface must include all of them. The singularity analysis is quite simple: we can easily see that none of the first or second singularities are spurious and therefore the three-dimensional surface is not required to contain any three- or two-dimensional objects. However, it is easy to see from the geometry that singularities of the form 
\begin{equation}
\la Y\,i\,i\pl1\ra = \la Y\,j\,j\pl1\ra = \la Y\,k\,k\pl1\ra = 0 \qquad\mbox{for $|i-j|,|j-k|,|k-i|>1$} \label{Triple1}
\end{equation}
are spurious (with the inequalities interpreted in a cyclic sense, in an obvious manner). There is also a spurious singularity when two indices are adjacent, for example if $j=i\pl1$. In that case there are two solutions for $Y$. Either $Y$ passes through $Z_{i\pl1}$ and intersects the line $(k\,k\pl1)$, or $Y$ is in a plane $(i\,i\pl1\,i\pl2)$ and intersects the same line. The first solution is allowed while the second is  spurious.
$$
\includegraphics[scale=0.6]{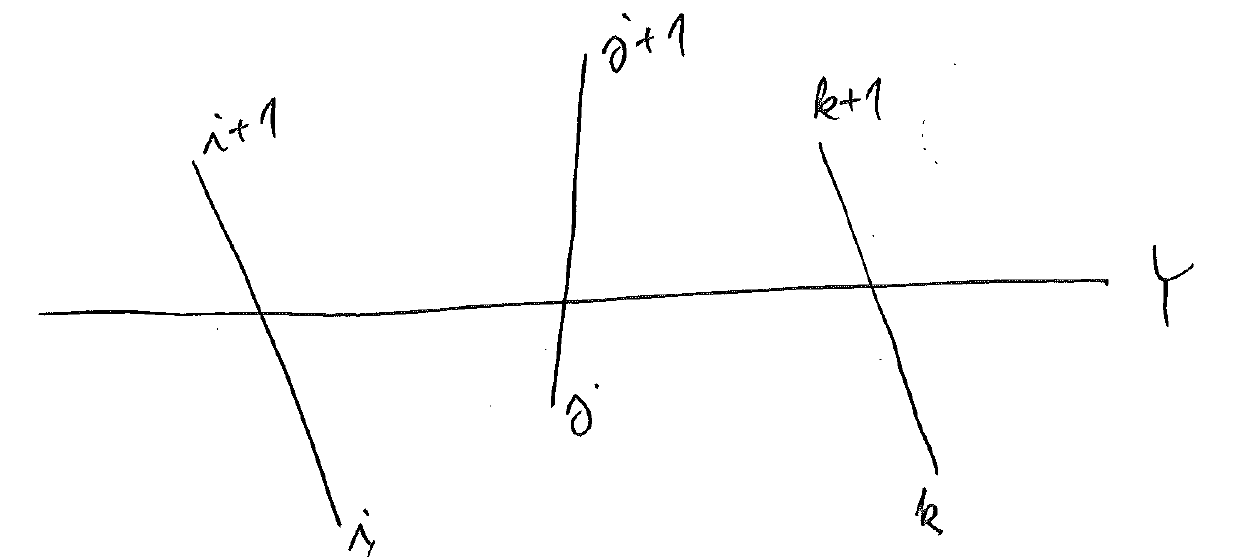}\qquad
\includegraphics[scale=0.6]{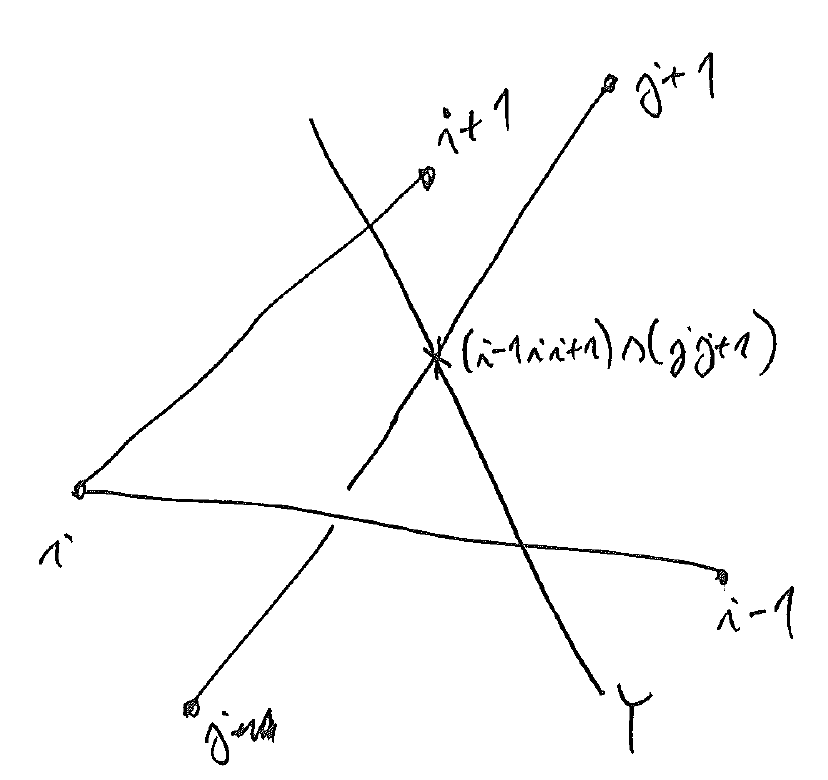}
$$
In the first case we can express $Y$ that satisfies (\ref{Triple1}) as
\begin{equation}
Y^\ast = (X\,j\,j\pl1)\cap(X\,k\,k\pl1) \equiv X_1 + \alpha X_2 + \alpha^2
X_3
\end{equation}
where $X=Z_{i}+\alpha Z_{i\pl1}$.
Importantly, $Y^\ast$ is quadratic in the parameter $\alpha$ which parametrizes the position
of $X$ on $(Z_iZ_{i\pl1})$. The numerator is a degree $(n-4)$
function of the lines $Y^\ast$, so after expansion it is a
polynomial of degree $2n-8$ in $\alpha$:
\begin{align}
{\cal N}(Y^\ast) = C\cdot (X_1+\alpha X_2 + \alpha^2 X_3)\dots (X_1+\alpha X_2 +
\alpha^2 X_3) = \sum_{k=0}^{2n-8} C^{(k)} \alpha^k = 0
\end{align}
where $C^{(k)}$ are contractions of $C$ with $X_1$, $X_2$, $X_3$.
This relation must be true for all $\alpha$ so we require all $2n-7$
coefficients $C^{(k)}$ to vanish.

For the second class of spurious singularities we have
\begin{equation}
Y^\ast=(X\,i\mi1\,i)\cap(X\,j\,j\pl1) \equiv X_1 + \alpha X_2
\end{equation}
where $X=Z_{i}+\alpha Z_{i\pl1}$.
We see that $Y^\ast$ is only
linear in $\alpha$, and ${\cal N}(Y^\ast)=0$ represents  $n-3$
constraints.

The spurious fourth singularities are easier to analyze as there are only a small set of legal singularities,
\begin{equation}
\la Y\,i-1\,i\ra=\la Y\,i\,i\pl1\ra = \la Y\,j\mi1\,j\ra = \la Y\,j\,j\pl1\ra = 0 \label{ForthAllowed}
\end{equation}
from which only the first solution $Y^\ast=(ij)$ is allowed. The second solution $Y^\ast=(i\mi1\,i\,i\pl1)\cap(j\mi1\,j\,j\pl1)$ as well as all other cases than (\ref{ForthAllowed}) are spurious and the numerator must kill them. Here $Y^\ast$ is fully specified so ${\cal N}(Y^\ast)=0$ is only a single constraint.

We have listed all the conditions on the numerator arising from spurious singularities, but not all the conditions are independent; in fact there is a large overlap between them.
We can make choices for which independent sets to take.  It is especially convenient to formulate the final list in terms of conditions that the numerator must vanish on certain points $Y^\ast$ rather than lines. The vanishing of the numerator on a line is then trivially implied if it also vanishes on sufficient number of points lying on the line.

One such choice is as follows. We impose that the numerator vanishes on all spurious fourth singularities except one set --- for each $i,j,k$ we omit one spurious point $Y^\ast$ for which
\begin{equation}
\la Y\,i\,i\pl1\ra = \la Y\,j\,j\pl1\ra = \la Y\,k\,k\pl1\ra = \la Y\,\ell\,\ell\pl1\ra =  0
\end{equation}
for some $\ell$ of our choice. But we have to be careful and do not choose the same point $Y^\ast$ multiple times (e.g.\ fix $i,j,\ell$ and choose $k$). It is easy to see that the total number of all spurious fourth singularities is $\frac{n(n-3)(n-4)(n+7)}{12}$ and the number of points we omit is $\frac{n(n-4)(n-5)}{6}$. The total number of constraints is then
\begin{equation}
\frac{n(n-3)(n-4)(n+7)}{12} - \frac{n(n-4)(n-5)}{6} = 2\left(\begin{array}{c}n\\4\end{array}\right)-\left(\begin{array}{c}n
\\3\end{array}\right) - 1
\end{equation}
matching exactly with (\ref{NumD}) up to one parameter, which is the overall constant.

It is interesting that we have completely determined the form from considerations of the ``ordinary" singularities, without separately considering the ``composite" singularities where a fourth boundary is reached by putting only three factors in the denominator to zero. These singularities are automatically matched correctly by our construction. As with the polygons, this formulation of the 1-loop MHV integrand is not associated with any ``triangulation" --- neither a BCFW nor a ``local" expansion --- but is instead directly determined by a complete picture of the amplituhedron geometry.

\subsection{Polytope in $\mathbb{P}^3$}

Let us now discuss the first non-trivial case of tree-level amplituhedron beyond $m=2$ kinematics which is for $m=3$, $k=1$  and $n=5$. It concerns the geometry of the polytope with five vertices in $\mathbb{P}^3$. This familiar object was discussed at length in \cite{Hodges:2009hk,ArkaniHamed:2010gg}. The explicit formula was found using both BCFW and local triangulation. This polytope is not cyclically invariant since $m$ is odd.  In order to make our discussion compatible with \cite{Hodges:2009hk,ArkaniHamed:2010gg} we choose to label the external points as $(13456)$, thus making $1$ and $3$ special and omitting label $2$. With this choice the final result directly corresponds to the 6-point NMHV split helicity amplitude $1^-2^-3^-4^+5^+6^+$.

The structure of the form is
\begin{equation}
\Omega_6 = \frac{\la Yd^3Y\ra\,{\cal N}(Y)}{\la Y134\ra\la Y145\ra\la Y156\ra\la Y345\ra\la Y356\ra\la Y136\ra}\label{P3form}
\end{equation}
where the poles consist of all the $\la Y\,i\,i\pl1\,j\,j\pl1\ra$, but omitting label $2$. The numerator ${\cal N}(Y) = C_{IJ}Y^IY^J \equiv (C\cdot YY)$ represents a quadric in $\mathbb{P}^3$. In total it has $d=10$ degrees of freedom and so can be specified, up to an overall constant, by nine equations of the type ${\cal N}(Y^\ast)=0$.

By a procedure similar to that of the previous section we can list all spurious singularities generated by the denominator which are absent in the numerator. The space is three-dimensional and therefore we must consider the second and third spurious singularities generated by the denominator. The only allowed second singularities are lines $(ij)$ and allowed third singularities are points $Z_k$. This gives us a list of six ``illegal'' lines $L_j$ which are {\em not} of this type:
\begin{align}
&L_1 = (134)\cap(156),\quad L_2=(345)\cap(136),\quad L_3=(145)\cap(356)\nonumber\\
&L_4=(145)\cap(136),\quad L_5 = (134)\cap(356),\quad L_6 = (156)\cap(345) \, .
\end{align}
The numerator then must vanish for all $Y^\ast = L_j^{(1)} + \alpha L_j^{(2)}$ where $L_j^{(1)}$, $L_j^{(2)}$ are two arbitrary points on the line $L_j$. That is,
\begin{equation}
{\cal N}(Y^\ast) = C_{IJ}(L_j^{I(1)} + \alpha L_j^{I(2)})(L_j^{J(1)} + \alpha L_j^{J(2)}) = C^{(0)}_j + \alpha C^{(1)}_j + \alpha^2 C^{(2)}_j = 0\, .
\end{equation}
This must be true for any value of $\alpha$, and hence we have three constraints for each line. 
There are also six spurious third boundaries, which are points $X_i$ not coincident with one of the $Z_j$, namely:
\begin{align}
&X_1 =(136)\cap(145)\cap(345),\quad X_2 =
 (136)\cap(145)\cap(356),\quad X_3= (134)\cap(156)\cap(356)\nonumber\\
&X_4 =(134)\cap(145)\cap(356),\quad X_5 = (134)\cap(156)\cap(345) ,\quad X_6=(136)\cap(156)\cap(345)
\end{align}\, .
$$
\includegraphics[scale=0.6]{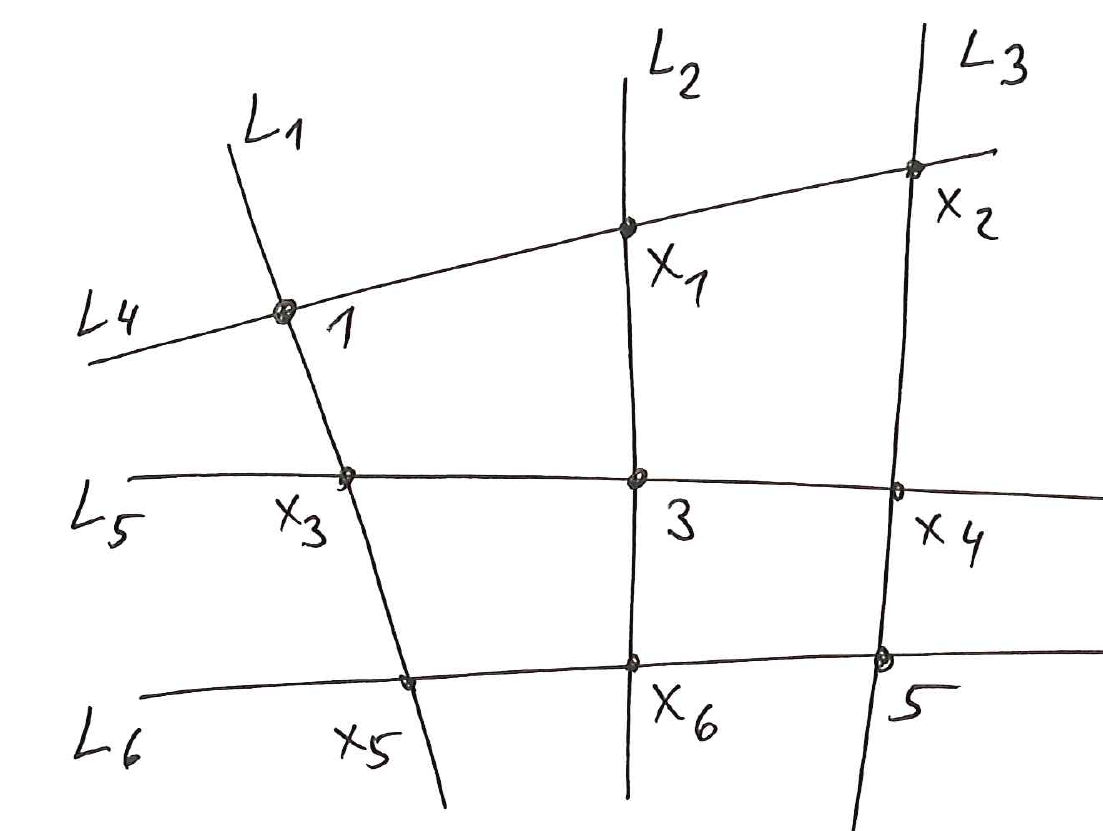}\label{P3points}
$$
On these, the numerator must simply vanish: ${\cal N}(Y^\ast=X_j)=0$. This looks very similar to the $\mathbb{P}^2$ case but there is a new phenomenon here. In $m=3$ kinematics the singularities for a generic numerator in (\ref{P3form}) are no longer logarithmic, because double poles are generated. We can see this explicitly when approaching the singularity $Y=Z_1$. We can first set $\la Y134\ra=\la Y145\ra=0$ by moving $Y$ on a line $(14)$, i.e.\ $Y=Z_1+\alpha Z_4$. In this case $\la Y156\ra = -\alpha\la 1456\ra$ and $\la Y136\ra = -\alpha \la1346\ra$ and the double pole in $\alpha$ is generated. The same phenomenon happens when we try to localize $Y$ to $Z_3$ and $Z_5$. If there are to be only logarithmic singularities, we must also require ${\cal N}(Y=Z_1) = {\cal N}(Y=Z_3) = {\cal N}(Y=Z_5)=0$. Putting these conditions together, it seems that there are too many constraints: six lines and nine points. But quite beautifully, we can easily see from the picture above that all the spurious points and lines are aligned so that all the required zeros of ${\cal N}$ are indeed possible.

The vanishing of the numerator on a line requires three constraints. In other words, if the numerator vanishes on three points on a given line, then it automatically vanishes on {\em all} points of this line. As can be seen easily from the picture, the numerator vanishes on all six spurious lines if we impose its vanishing on all nine points, ${\cal N}(Y^\ast)=0$ for $Y^\ast = X_1,\dots X_6, Z_1,Z_3,Z_5$. This imposes exactly the nine constraints necessary to fix the numerator completely (up to an overall constant). This means that the zero surface for this amplituhedron is the quadric in $\mathbb{P}^3$ specified by these nine points, which mostly lies outside the polytope, touching it on three points $Z_1$, $Z_3$ and $Z_5$.

\subsection{Polytope in $\mathbb{P}^4$}

Let us now consider the same exercise for $m=4$ kinematics. This is a cyclic case and for $n=6$ (and $k=1$) the form is
\begin{equation}
\Omega = \frac{\la Y\,d^4Y\ra\,{\cal N}(Y)}{\la Y1234\ra\la Y1245\ra\la Y1256\ra\la Y2345\ra\la Y2356\ra\la Y2361\ra\la Y3456\ra\la Y3461\ra\la Y4561\ra}\label{P4form}
\end{equation}
where the numerator is quartic, ${\cal N}(Y) = C_{IJKL}Y^IY^JY^KY^L$, and it has 69 degrees of freedom (up to overall scale). We can repeat the same exercise from the previous subsection by demanding that the numerator vanishes on all spurious singularities. The space is four-dimensional and the only legal boundaries are planes $(i\, i\pl1\, j)$, lines $(ij)$ and points $(i)$. The first spurious singularities are then the planes defined by two terms in the denominator being set to zero, e.g.\ $P =(1234)\cap(1256)$. An arbitrary point $Y$ on this plane can be described using three parameters
$\alpha_1$, $\alpha_2$, $\alpha_3$ as
\begin{equation}
Y^\ast = \alpha_1 Z_1 + \alpha_2 Z_2 + \alpha_3 (Z_3 - \sigma^\ast Z_4)\qquad \mbox{where}\quad \sigma^\ast = \frac{\la 12356\ra}{\la 12456\ra}\, .
\end{equation}
Plugging into the numerator we get
\begin{equation}
{\cal N}(Y^\ast) =  C_1 \alpha_1^4 + C_2 \alpha_1^3\alpha_2 + C_3 \alpha_1^3\alpha_3 + \dots C_{15}\alpha_3^4 = 0
\end{equation}
where the $C_i$ are some independent constants. We demand that all $C_i$ must vanish, and this imposes 15 constraints. Similarly, we demand that the numerator vanishes on all spurious lines, for example $L = (1234)\cap(1256)\cap(2345)$. We now need two parameters,
\begin{equation}
Y^\ast = \alpha_2 Z_2 + \alpha_3 (Z_3 - \sigma^\ast Z_4)
\end{equation}
Plugging into the numerator we get
\begin{equation}
{\cal N}(Y^\ast) = D_1 \alpha_2^4 + D_2 \alpha_2^3\alpha_3 + D_3 \alpha_2^2\alpha_3^2 + D_4 \alpha_2\alpha_3^3 + D_5 \alpha_3^4 = 0
\end{equation}
and vanishing on the line imposes the 5 constraints that all $D_i$ must vanish. Finally, the numerator must vanish on all spurious points which are not $Z_i$, for example $X=(1234)\cap(3456)\cap(1256)\cap(2345)$. It is easy to write a list of spurious planes, lines and points which are not among legal boundaries but the form generates a singularity
when we place $Y$ on them. The list is quite long and contains 18
planes, 42 lines and 45 points. We can list conditions for all these
illegal configurations but they are not independent and the overlap
is substantial.

We will shortly give an independent set of constraints, but before doing that we have to deal with the second class of constraints which come from the demand that  double poles are cancelled by the numerator. This is similar to the previous subsection but with the extra complication that the double poles can be generated when we move $Y$ to a particular line, i.e.\ to a third boundary rather than to a point. For example, we can set $\la Y1234\ra = \la Y1245\ra = 0$ and then $Y=Z_1+\alpha Z_2 + \beta Z_4$. The other two terms then produce $\la Y1256\ra = \beta \la 12456\ra$, $\la Y1236\ra = -\beta \la 12346\ra$ and the form (\ref{P4form}) generates a double pole in $\beta$. This means that effectively the line $(12)$ is spurious, and the numerator must vanish when $Y$ is put on it. Similarly for line $(14)$ and all cyclically related cases.
If we further localize $Y$ to the point $Z_1$, we encounter possible {\em triple} poles in the form, which must be cancelled by the numerator.
For example, setting $\la Y1234\ra = \la Y1236\ra = \la Y1346\ra =0$ by localizing $Y$ on a line $(13)$ (which is not required to be cancelled by the numerator), puts  $Y=Z_1 + \alpha Z_3$. Now the three remaining poles $\la Y1245\ra$, $\la Y1256\ra$, $\la Y1456\ra$ produce a factor of $\alpha^3$ in the denominator. The numerator must therefore vanish quadratically with $\alpha$ to kill this triple pole, which requires
\begin{equation}
(C\cdot Z_1Z_1Z_1X) = 0\qquad \mbox{for\,\,arbitrary\,\,$X$}
\end{equation}
and similarly for all other $Z_j$. This list of constraints must be combined with the previous one.

We can now proceed to choosing an independent set from this list of constraints, which give exactly the correct number of linear equations to fix the numerator. There are many choices that work, but it is again convenient to give vanishing constraints on enough individual points as to automatically enforce the needed vanishing conditions on lines and planes.  One choice is the following:

\begin{itemize}
\item We demand the numerator vanishes on illegal points:
\begin{equation}
(C\cdot YYYY) = 0 \quad \mbox{for
$Y=(1234)\cap(3456)\cap(1256)\cap(X)$}
\end{equation}
where $X$ may be $(1245), (2356), (3461)$. This gives 3 constraints.

\item We demand the vanishing of the numerator on illegal points on the line $(12)$, i.e.\ on
$Y=(12)\cap(3456)$, like
\begin{equation}
(C\cdot YYYX)=0 \qquad\mbox{where\,\,$X=\alpha_3 Z_3 + \alpha_4 Z_4 +
\alpha_5 Z_5 + \alpha_6 Z_6$}
\end{equation}
for arbitrary $\alpha_3,\alpha_4,\alpha_5,\alpha_6$. Similarly for
points on lines
$(23)$, $(34)$, $(45)$, $(56)$, $(61)$, $(14)$, $(25)$, $(36)$. Each case
gives 4, in total 36 constraints.

\item Finally we demand that the numerator vanishes for all points
$Z_i$ like
\begin{equation}
(C\cdot Z_iZ_iZ_iX)=0\qquad\mbox{for\,\,arbitrary $X$}\, .
\end{equation}
Each point $Z_i$ gives 5 constraints (the number of degrees of
freedom in generic $X$), and so 30 constraints.

\end{itemize}
The total number of constraints is $3+36+30=69$, which is, beautifully,  the correct number to fix the degrees of freedom of the numerator.

We can finally repeat the same exercise for arbitrary $n$. The form $\Omega$ is
\begin{equation}
\Omega = \frac{\la Y\,dY\,dY\,dY\,dY\ra\, {\cal N}(Y,Z_i)}{\prod_{i,j}\la i\,i\pl1\,j\,j\pl1\ra}
\end{equation}
The only legal singularities beyond $(i\,i\pl1\,j\,j\pl1)$ are planes $(i\,i\pl1\,j)$, lines $(i\,j)$ and points $(i)$. Everything else is spurious. The numerator has degree $m=\frac{n(n-3)}{2}-5$ in $Y$:
\begin{equation}
{\cal N} = C_{I_1I_2\dots I_m}Y^{I_1}Y^{I_2}\dots Y^{I_m}\equiv
(C\cdot YY\dots Y)\, ,
\end{equation}
where $C$ is a symmetric tensor in all $m$ indices. The total number of degrees of freedom is
\begin{equation}
d = \left(\begin{array}{c}m+4 \\ 4
\\\end{array}\right)-1 = \left(\begin{array}{c}\frac{n(n-3)}{2}-1 \\ 4
\\\end{array}\right)-1
\end{equation}
plus one for an overall constant. The analysis is very similar to the six-point case; the only difference is that multiple poles are now generated when $Y$ approaches lines $(i\,i\pl1)$ or points $(i)$. We will not present the details of the analysis here,  but rather provide the full list of vanishing constraints on ${\cal N}(Y)$. The first set of conditions is from lines:
\begin{itemize}
\item $(C\cdot YY\dots Y\underbrace{XX\dots X}_{n-6})=0$ for $Y$ on the line $(i\,i\pl1)$.
\item No condition on the lines $(i\,i\pl2)$, ie. $(13)$, $(24)$, $\dots$, $(n2)$.
\item $(C\cdot YY\dots Y)=0$ for $Y$ on all other lines $(ij)$.
\end{itemize}

The second set comes from localizing $Y$ into special points
\begin{itemize}
\item $(C\cdot YY\dots Y\underbrace{XX\dots X}_{2n-11})=0$ for $Y=Z_i$.
\item $(C\cdot YY\dots Y\underbrace{XX\dots X}_{n-5})=0$ for $Y=(i\,i\pl1)\cap(j\,j\pl1\,k\,k\pl1)$, for generic $j$, $k$. These are conditions on special points on line $Z_iZ_{i\pl1}$.
 \item $(C\cdot YY\dots YX)=0$ for $X=(i\,j)\cap(k\,k\pl1\,\ell\,\ell\pl1)$, for generic $k$, $\ell$. These are conditions on special points on line $Z_iZ_j$ (not $Z_iZ_{i\pl2}$).
 \end{itemize}

These conditions are redundant. Selecting the independent set (there exists a choice when only conditions on points are imposed) we find exactly the number $d$ of conditions needed to fix the numerator.

\subsection{Summary}

In this section we showed in four different examples how to construct the numerator for the form $\Omega$ which has an interpretation as a zero surface. It lies outside the amplituhedron space and contains all spurious points, lines and higher planes that occur as intersections of the local poles of $\Omega$. It also touches the amplituhedron space where iterated residues of $\Omega$ could generate multiple poles.

\begin{enumerate}
\item We first studied the simplest case of $k=1$ and $m=2$ where $\Omega$ generates spurious points only. The zero surface is then an algebraic curve which is defined by containing all $\frac{n(n-3)}{2}$ spurious points.
\item The next case was $k=2$ and $m=2$, where in addition to points there are also spurious lines. The zero surface is a 2-dimensional plane which contains all spurious points and lines. For $m=2$ and general $k$ the surface is a $k$-dimensional hyperplane which contains $k-2,k-3,\dots,1$-dimensional projective objects.
\item For $m=3$ kinematics we looked at the 5-point $k=1$ case which has a direct physical interpretation as a split-helicity amplitude. In that case the surface contains spurious points and lines, but is also  required to touch the positive space in three points $Z_1,Z_3,Z_5$. The reason is that at these positions $\Omega$ generates double poles which must be cancelled by the numerator.
\item Finally we looked at $m=4$ kinematics for $k=1$. Then in addition to spurious planes, lines and points we have points and lines on the boundaries of the positive space which the zero surface contains. But there is also a new phenomenon: the numerator is required to vanish more strongly than just ${\cal N}(Y^\ast)=0$ due to the presence of multiple poles in the denominator of $\Omega$, both for lines and points.
\end{enumerate}
With these examples we have covered all the qualitatively different constraints that must be imposed on the numerator for general $n$, $k$ and $m$ for the tree-level case.

\section{Further Checks of Positivity}

We expect that for general $n,k,L$, the numerator of the amplitude form can be completely fixed in a ``global" way by analogous considerations:  vanishing on spurious singularities and killing multiple poles.  If the ``bad" singularities indeed continue to lie ``outside" the amplituhedron as in the examples we have seen, the positivity of the form would follow. The geometry involved will however certainly become much more intricate beyond the simple examples we have already considered. In this section, therefore, we take a more ``experimental" tack, and give further direct evidence for the positivity of the form without a complete understanding of the geometry. We can do this most straightforwardly by numerically checking that the form is positive, or by finding an expansion of the amplitude form where the positivity is manifest term-by-term.

Of course, the latter approach is preferable, but none of the systematic expansions for amplitudes, based on BCFW recursion relations or MHV diagrams, make positivity manifest. For instance BCFW recursion relations \cite{Britto:2004ap,Britto:2005fq,ArkaniHamed:2010kv} have spurious poles which do not have uniform signs inside the amplituhedron --- obviously, to have any hope of being manifestly positive, the expansion must have only local poles.

Curiously, we have previously seen expansions of the amplitude with only local poles --- indeed the most compact expressions we have seen for loop level integrands were of this type. However, these ``local expansions" did not appear have an obvious conceptual purpose in life. For instance they do not actually make locality manifest, since (especially starting with NMHV amplitudes), individual terms have sets of poles that are mutually incompatible (analogous to simultaneously having $s$ and $t$ channel poles in four-particle scattering). Furthermore, while in the simplest cases these ``local forms" seemed quite canonical, at higher points and loops there seemed to be many ways of expressing them, so it was not clear what these forms were trying to tell us.

We now have a natural rationale for the existence of these ``local forms": while they don't make locality manifest, their purpose should be to make the positivity of the form manifest! As we will see, in precisely those cases where the local expansions are completely canonical, they {\it do} make the positivity of the form manifest term-by-term, although seeing this analytically requires certain inequalities that quite non-trivially follow from the positive structure of the amplituhedron. However in general, when the local forms are less canonical, we find that they don't make positivity of the form manifest term-by-term. Nonetheless all our numerical checks non-trivially verify positivity of the form. It would be extremely interesting to search for ``manifestly positive" representations of the amplitude form --- these are likely to be extremely canonical, and may give clues to the ``dual amplituhedron" picture we seek,
much as the BCFW expansion pointed to the amplituhedron itself.

\subsection{All $k$ for $m=2$ kinematics}

For $m=2$ kinematics we know the amplitude form explicitly for all $k$ and $n$, in two different triangulations. The space here is very simple and can roughly be characterized as ``(polygon)$^k$", and triangulations of the polygon lift directly to triangulations for the general case. 

The local triangulation for a polygon \cite{ArkaniHamed:2010gg} is
\begin{equation}
\Omega_n^{(1)} = \sum_i \frac{\la 12\,i\ra\la i\mi1\,i\,i\pl1\ra}{\la Y12\ra\la Y\,i\mi1\,i\ra\la Y\,i\,i\pl1\ra}
\end{equation}
where the positivity is manifest term-by-term. For $k=2$ we get
\begin{equation}
\Omega_n^{(2)} = \sum_{i,j} \frac{\la 12\,i\,j\ra\la Y\,(i\mi1\,i\,i\pl1)\cap(j\mi1\,j\,j\pl1)\ra}{\la Y12\ra\la Y\,i\mi1\,i\ra\la Y\,i\,i\pl1\ra\la Y\,j\mi1\,j\ra\la Y\,j\,j\pl1\ra}
\end{equation}
where we use the notation $\la Y\,(i\mi1\,i\,i\pl1)\cap(j\mi1\,j\,j\pl1)\ra = \la Y_1\,i\mi1\,i\,i\pl1\ra\la Y_2\,j\mi1\,j\,j\pl1\ra - \la Y_2\,i\mi1\,i\,i\pl1\ra\la Y_1\,j\mi1\,j\,j\pl1\ra$
and this case also corresponds to the integrand of MHV 1-loop amplitude for $m=4$ kinematics once we identify line $Y$ with the loop. This is not manifestly positive, due to the second term in the numerator. Expanding $Y$ in some basis it is obvious that the positivity of the full expression relies on
\begin{equation}
\la a\,b\,(c\mi1\,c\,c\pl1)\cap(d\mi1\,d\,d\pl1)\ra>0\qquad\mbox{for $a<b<c<d$}\, .
\end{equation}
This simple statement does not trivially follow from the positivity of the all the ordered minors $\langle i j k l \rangle > 0$ for $i<j<k<l$. A relatively simple inductive proof begins by recalling an important general fact about positive grassmannians: we can build any positive configuration for the external $Z$'s starting from a zero-dimensional cell in $Z$-space --- where all but 4 $Z$'s are set to zero --- and building the general configuration by successively (positively) shifting columns by their neighbors. It is trivial to see that the above expression is positive for $Z$'s in the zero-dimensional cells, and a small computation shows that they can only increase under the action of bridges. But while this argument can be used to probe positivity, it doesn't give any insight into why we might have even imagined this object was positive to begin with. It would be very nice to have a more conceptual proof of this surprising fact, since much more intricate analogs of this statement will be true for higher $k$ and loop orders. At least for for $k=2$, \cite{Lam:2014jda} appears to provide a deeper explanation, where our expression is part of a ``canonical basis" of positive objects built out of minors.

Moving on, we have found the triangulation of the $m=2$ amplituhedron for general $k$ and determined the corresponding form, which turns out to be
\begin{equation}
\Omega^{(k)}_n = \sum_{j_1,\dots j_k}\frac{\la 12\,j_1\,j_2\dots\,j_k\ra\la Y\,(j_1\mi1\,j_1\,j_1\pl1)\cap (j_2\mi1\,j_2\,j_2\pl1)\dots (j_k\mi1\,j_k\,j_k\pl1)\ra}{\la Y12\ra\la Y\,j_1\mi1\,j_1\ra\la Y\,j_1\,j_1\pl1\ra\la Y\,j_2\mi1\,j_2\ra\la Y\,j_2\,j_2\pl1\ra \dots \la Y\,j_k\mi1\,j_k\ra\la Y\,j_k\,j_k\pl1\ra}
\end{equation}
for
\begin{equation}
\la Y\dots\ra = \sum_\sigma (-1)^\sigma \prod_{p=1}^k\la \hat{Y}_{\sigma(p)}\,j_p\mi1\,j_p\,j_{p\pl1}\ra \label{knum}
\end{equation}
where we define $\hat{Y}_p=(Y_1\dots Y_{p\mi1}Y_p\dots Y_k)$ and we sum over all permutations $\sigma$. The positivity of (\ref{knum}) is again not obvious but can be proved inductively along the same lines as for $k=2$.

\subsection{Tree amplitudes}

We now consider the tree amplitudes associated with the tree amplituhedron for $m=4$. For the case $k=1$, we have both the volume picture and our general understanding of the numerator, which guarantee positivity. We can also see the positivity directly from the local triangulation found in \cite{ArkaniHamed:2010gg}, which gives the amplitude as a sum of manifestly positive terms:
\begin{equation}
\Omega_n^{(1)} =
\sum_{j,i,s} \frac{\la 1234\,j\ra\la
j\mi1\,j\,j\pl1\,j\pl2\,i\ra\la j\mi1\,j\,j\pl1\,i\,i\mi s\ra\la
j\,j\pl s\,i\mi1\,i\,i\pl1\ra}{\la Y1234\ra\la
 Y\,j\mi1\,j\,j\pl1\,j\pl2\ra\la
Y \,j\mi1\,j\,i\mi1\,i\ra\la Y\,j\,j\pl1\,i\,i\pl1\ra\la Y\,j\,j\pl
s\,i\,i\mi s\ra}
\end{equation}
where $j=1,\dots,n$, $i=j\pl2,\dots,j\mi2$ (in cyclic sense) and $s=\pm1$. The choice of the reference plane $(1234)$ is arbitrary and we can choose any other one.

For higher $k$ we do not have such an expansion and so we can only test positivity numerically. We know all the tree amplitudes by using BCFW recursion relations, conveniently available in the Mathematica packages \cite{Bourjaily:2010wh,Bourjaily:2012gy}, as well as in CSW expansion written in nice closed forms in momentum twistor space \cite{CSW}. All terms in both expansions can easily be uplifted to the $Y$-space of the amplituhedron. Thus we can evaluate the amplitudes for a huge set of points inside the amplituhedron, and see if the result is numerically positive. We did this check using both BCFW and CSW expansions for N$^2$MHV up to $n=12$ for $10^4$ points each, both in inside the positive space and near the boundary. Again, the positivity looks very non-trivial in both expansions, with huge cancellations between positive and negative terms leaving a positive result.
As an example of an explicit formula in the $Y$-space, we write all N$^2$MHV amplitudes in the CSW expansion (here $Y$ is a line),
\begin{equation}
\Omega_n^{(2)} = \hspace{-0.25cm}\sum_{i<j\leq k<\ell} \frac{\la
Y\,(X\,i\,i\pl1\,j\,j\pl1)\cap(X\,k\,k\pl1\,\ell\,\ell\pl1)\ra^4}{\begin{array}{c}\la
Y\,X\,i\,i\pl1\,j\ra\la Y\,X\,i\,i\pl1\,j\pl1\ra\la Y
X\,i\,j\,j\pl1\ra\la Y\,i\pl1\,j\,j\pl1\ra\la
Y\,i\,i\pl1\,j\,j\pl1\ra\\
\la Y\,X\,k\,k\pl1\,\ell\ra\la Y\,X\,k\,k\pl1\,\ell\pl1\ra\la Y
X\,k\,\ell\,\ell\pl1\ra\la Y\,k\pl1\,\ell\,\ell\pl1\ra\la
Y\,k\,k\pl1\,\ell\,\ell\pl1\ra\end{array}}\, .
\end{equation}
Here the boundary term $j=k$ has a special form, as two of the factors in the denominator get entangled in the quadratic pole in $Y$:
$$
\hspace{-0.3cm}\frac{\la
Y\,(X\,i\,i\pl1\,j\,j\pl1)\cap(X\,j\,j\pl1\,\ell\,\ell\pl1)\ra^4}{\begin{array}{c}\la
Y X\,i\,j\,j\pl1\ra\la
Y\,i\pl1\,j\,j\pl1\ra\left[\la Y\,X\,i\,i\pl1\,j\pl1\ra\la
Y\,X\,\ell\,\ell\pl1\,j\ra - \la Y\,X\,i\,i\pl1\,j\ra\la
Y\,X\,\ell\,\ell\pl1\,j\pl1\ra\right]\\ \la
Y\,i\,i\pl1\,j\,j\pl1\ra\la Y\,X\,i\,i\pl1\,j\ra \la
Y\,X\,j\,j\pl1\,\ell\ra\la Y X\,j\,j\pl1\,\ell\pl1\ra\la
Y\,\ell\,\ell\pl1\,j\pl1\ra\la
Y\,j\,j\pl1\,\ell\,\ell\pl1\ra\end{array}}\, .
$$
The brackets in the numerator have the meaning of
$$
\la Y(a_1b_1c_1d_1e_1)\cap(a_2b_2c_2d_2e_2)\ra = \la Y_1 a_1b_1c_1d_1e_1\ra\la Y_2a_2b_2c_2d_2e_2\ra - (Y_1\leftrightarrow Y_2)
$$
where $Y_1$, $Y_2$ are two points on a line $Y$. Unlike the BCFW expansion which triangulates the amplituhedron internally, this is not true for CSW. Not only can the generic $X$ lie outside the space, but there even seems to be no choice for $X$ such that it is inside \cite{Lionel}.

In the $k=1$ case the local expansion is directly a triangulation of the dual amplituhedron, and therefore it is term-by-term positive. This is what we mean by canonical representation. For $k>1$ we do not have any local expansion; we suspect that if one is found it would give us a much better idea about what the ``dual amplituhedron" might be.   However, there is an interesting piece of data for $k>1$ which is encouraging for the existence of such forms and dual amplituhedra. The idea is simple: we want to repeat the exercise that we did for $m=2$ to get the $k=2$ case from the $k=1$ case --- entangling two copies of $k=1$ amplitudes --- but now for higher $k$. Let us start with $m=3$, for which the local triangulation was found in \cite{Hodges:2009hk,ArkaniHamed:2010gg}, for the case of split helicity $1^-2^-3^-4^+\dots n^+$ amplitude:
\begin{equation}
A_n^{(2)} = \sum_{i=5}^{n}\sum_{s=\pm1} \frac{\la Yd^3Y\ra \la 134\,i\ra\la 2\pl
s\,i\mi1\,i\,i\pl1\ra\la13\,i\mi
s\,i\ra}{\la Y134\ra\la Y1\,i\mi1\,i\ra\la Y3\,i\,i\pl1\ra\la Y\,2\pl
s\,i\,i\mi s\ra}\label{split}\, .
\end{equation}
Consider a similar expression $A_n^{(3)}$ with the origin $3$, and so corresponding to the $1^+2^-3^-4^-5^+\dots n^+$ amplitude. Now we put indices $2$ and $3$ back and write an expression which is formally a $k=2$, $m=4$ tree-type amplitude:
\begin{equation*}
\hspace{-0.65cm}{\cal A}_n = \hspace{-0.5cm}\sum_{\begin{array}{c} \scriptstyle i=j\dots n \vspace{-0.2cm}\\
\scriptstyle j=5\dots n\mi1\vspace{-0.2cm}\\\scriptstyle
s_1,s_2=\pm1\end{array}} \hspace{-0.45cm} \frac{d\mu\,\,\la 1234ij\ra \la Y (2\,2\pl s_1\,i\mi1\,i\,i\pl1)\cap(3\,3\pl s_2\,j\mi1\,j\,j\pl1)\ra \la Y (123\,i\mi s_1\,i)\cap(234\,j\mi s_2\,j)\ra}{\la Y1234\ra\la Y12\,i\mi1\,i\ra\la Y23\,i\,i\pl1\ra\la Y2\,2\pl s_1\,i\,i\mi s_1\ra\la Y23\,j\mi1\,j\ra\la Y34\,j\,j\pl1\ra\la Y 3\,3\pl s_2\,j\,j\mi s_2\ra}
\end{equation*}
where $d\mu = \la Yd^4Y_1\ra\la Yd^4Y_2\ra$ and $Y$ is a line in $\mathbb{P}^5$. Note that this expression is not projective in $Y$ and $Z_2$, $Z_3$ and therefore does not qualify to be a candidate to be a proper $m=4$, $k=2$ formula. But even though it isn't an amplitude form, it isn't a random expression either: we can extract a physical amplitude out of it! Indeed the N$^2$MHV split helicity $1^-2^-3^-4^-5^+6^+\dots n^+$ amplitude can be found by 
using the standard procedure of extracting $Z,\eta$ out of six dimensional ${\cal Z}$ \cite{Arkani-Hamed:2013jha}, and integrating over $d^4\eta_2\,d^4\eta_3$. This gives a ``local expansion" at least for split-helicity amplitudes. And quite nicely, each term in this expansion is positive when evaluated inside the $m=4,k=2$ amplituhedron. This can be generalized to split-helicity amplitudes for all $k$ \cite{JaraNotes}.

\subsection{Loop integrands}

Let us move on to the loop integrand. The simplest case ($k=0$) of the $L=1$ loop MHV integrand has already been discussed, so we start with
$L=2$. The two loop variables correspond to two lines $AB$ and $CD$ in momentum twistor space. The positivity rules dictate that the expansion coefficients of $AB$ and $CD$ in terms of external $Z_i$ are positive matrices $G_+(2,n)$, while all $(4\times4)$ minors of the combined $(4\times n)$ matrix are also positive. The local expansion found in \cite{ArkaniHamed:2010kv} gives the amplitudes as a sum for $i<j<k<\ell$ over double pentagons
$$
\includegraphics[scale=0.6]{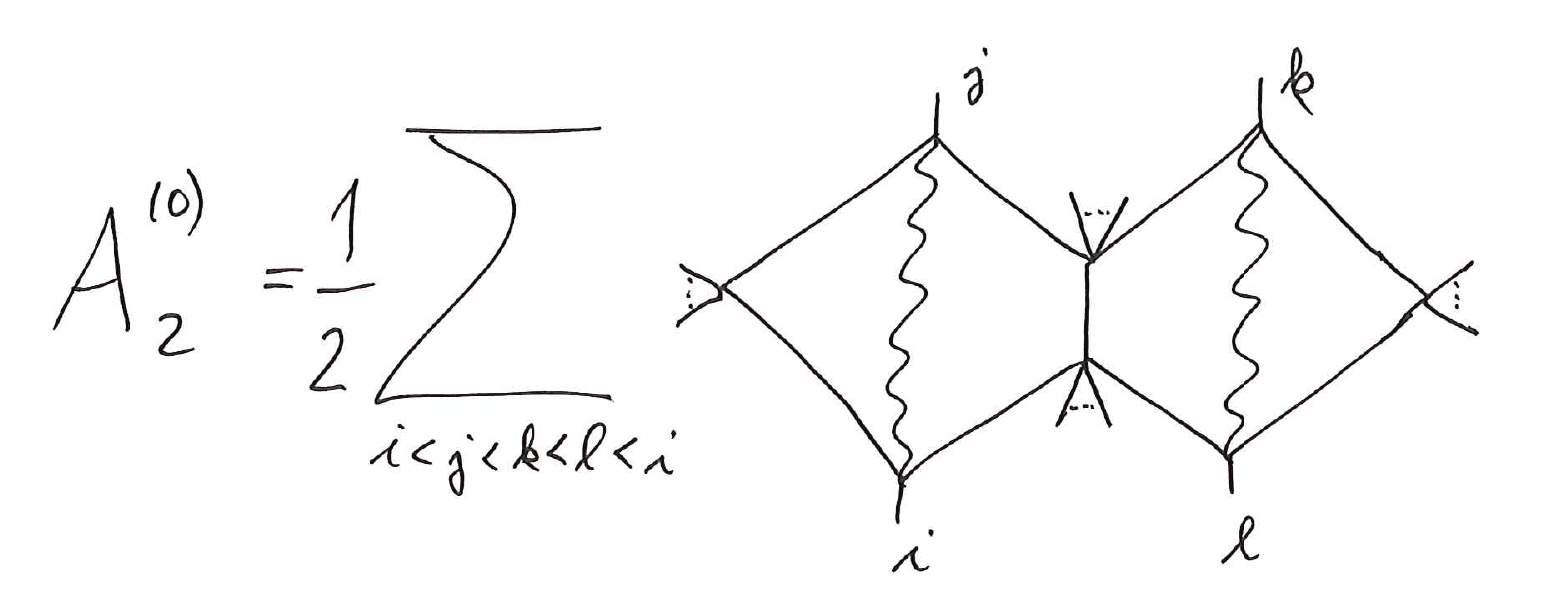}
$$
where the explicit expression for the double pentagon $Q_{ijkl}$ is
\begin{equation}
\hspace{-0.8cm}\frac{\la ijk\ell\ra\la AB\,(i\mi1\,i\,i\pl1)\cap(j\mi1\,j\,j\pl1)\ra\la CD\,(k\mi1\,k\,k\pl1)\cap(\ell\mi1\,\ell\,\ell\pl1)\ra}{\la AB\,i\mi1\,i\ra\la AB\,i\,i\pl1\ra\la AB\,j\mi1\,j\ra\la AB\,j\,j\pl1\ra\la ABCD\ra\la CD\,k\mi1\,k\ra\la CD\,k\,k\pl1\ra\la CD\,\ell\mi1\,\ell\ra\la CD\,\ell\,\ell\pl1\ra}\label{DoublePent}
\end{equation}
The numerator is manifestly positive because it is just two copies of the $L=1$ numerator which we discussed before. The denominator is also manifestly positive, including the term $\la ABCD\ra > 0$ for which the positivity of all $(4\times4)$ minors matter. 
 
 We also have an explicit result for $L=3$ for any $n$ given in \cite{ArkaniHamed:2010gh}. In this case the expansion is not positive term-by-term and we only perform a numerical check. We tested for more than $10^5$ points in different parts of kinematical regions that indeed the expression is positive. The discussion simplifies when we restrict to $n=4$. First, it is easy to show that there exist natural local building blocks which are manifestly positive in the positive region. These are nothing than standard scalar integrals used in the literature in the context of unitary methods. Second, there are reference data in the literature up to $L=7$ for which we can test the conjecture. For $L=3$ the amplitude \cite{Bern:2005iz} is a sum of two different building blocks (plus terms related by symmetry) with $+1$ coefficients, which makes the positivity of the final result completely manifest. Starting at $L=4$ we start to have both plus and minus coefficients. The final result is a sum of eight terms \cite{Bern:2006ew}
$$
\includegraphics[scale=0.6]{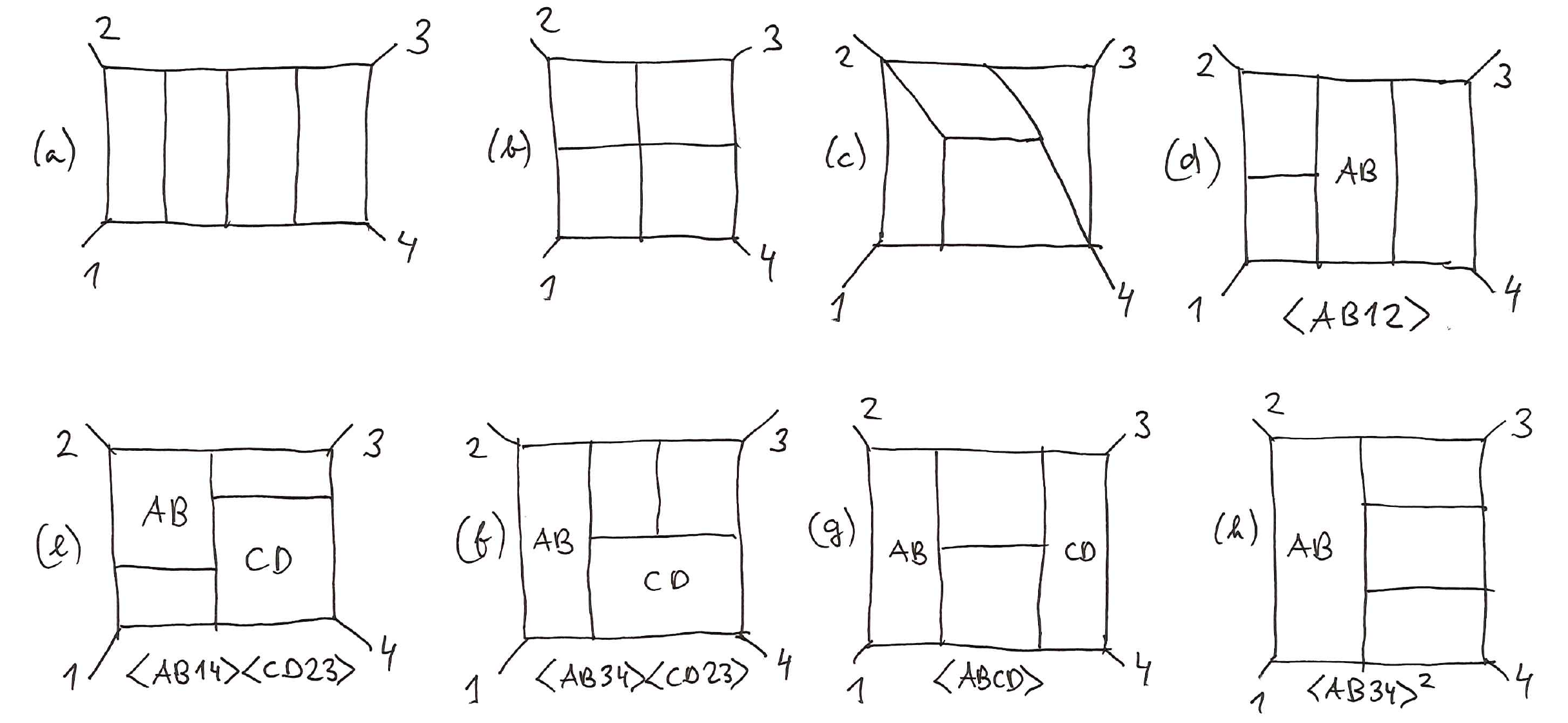}
$$
where the denominator is given by the propagators in the diagrams and we have omitted the constant $\la1234\ra$ factors in the numerator. It is easy to see that the expressions for all diagrams are individually positive when evaluating in the positive space. The coefficient in the expansion for the amplitude are $+1$ for integrals (a), (d)-(h) while for (b) and (c) we get $-1$ with proper symmetrization in loop momenta as well as cyclic sum over external legs. There is no obvious way to rewrite this as a sum of positive terms. However, the numerical checks confirmed that despite not being manifestly positive the full sum is in fact positive when evaluated in the positive region.

Going to higher $L$ the situation is more dramatic and the number of positive and negative terms in the result is almost $50:50$ in the end. The data are available up to 7 loops \cite{Bern:2005iz,Bern:2006ew,Bern:2007ct, Bourjaily:2011hi, Bern:2012di, Eden:2012tu}.
However, when evaluated numerically the final sum is always positive, which we checked up to $L=5$ for $10^4$ points for each loop order. We again checked the points inside the positive space and also near the boundary. Looking at the explicit numerical data, it indeed seems like a miracle that the sum of terms always stays positive as we are summing big positive and negative values, that always conspire to give a positive result.

Actually, there is a good reason why the expansion using scalar integrals does not make positivity manifest term-by-term. All terms are individually positive (and sum have negative coefficients in the amplitude) but they are in fact more positive than we need: they are positive in a bigger region than just the positive region given by the amplituhedron ${\cal A}_{n,k=0,L}$.

The existence of local expansion or MHV integrands is not so special as it was in the tree-level case for $k>0$. We can see it at the 4-loop 4-point example when the local expansion was not manifestly positive. The reason for that is that our building blocks --- local integrands --- are ``too positive" and perhaps we have to recombine them into ``less positive" building blocks \cite{Lam:2014jda} to make the positivity manifest. This would give us the canonical expansion which might then directly related to the triangulation picture for the dual amplituhedron.

We can also do the check for non-MHV amplitudes looking at the data available in the literature. The integrand for all one-loop amplitudes, $L=1$ for any $n$ and $k$, has been found in a ``local form" explicitly in \cite{Bourjaily:2013mma} (see earlier result \cite{Elvang:2009ya} for $k=1$) while for $L=2$ the only available class of results is for NMHV, ie. $k=1$, $L=2$ for any $n$, found in \cite{ArkaniHamed:2010gh}. Here, not even the ``local forms" can hope to make the positivity manifest,  as the result is always organized as the $A = (\mbox{Yangian invariant})\times (\mbox{integral})$. While the second part can be constructed to be positive, the Yangian invariant is never positive because of the presence of spurious poles. It would be very interesting to find such an expansion even for $k=1$ and $L=1$ where only simple R-invariants appear. This would force us to write a {\it super-local} expansion, both in external poles $\la Y\,i\,i\pl1\,j\,j\pl1\ra$ as well as internal poles $\la YAB\,i\,i\pl1\ra$, and perhaps see some interesting structures which would shed light on the origin of the positivity. But we have explicitly verified the positivity of the form numerically, for $L=1$: $k=1$ up to $n=12$ for $10^4$ points each, for $k=2$ up to $n=9$ for $10^3$ points each, and also for $L=2$, $k=1$ up to $n=7$ for $10^3$ points each. Again the points were chosen to be inside the space as well as near the boundary.

\subsection{Log of MHV Amplitude and the Ratio Function}

Interestingly, in addition to the positivity conjecture for the form for the amplitude form, we have also found similar statement to be true for two other standard and natural objects: the logarithm of the MHV amplitude and ratio function.

The integrand for scattering amplitudes has soft and collinear singularities; integrating the $L$-loop amplitude has a log$^2L$ infrared divergence as a consequence. However as is well-known, the IR divergences exponentiate, and it is natural to take the logarithm of the amplitude
\begin{equation}
{\cal A} = 1 + gA_1 + g^2A_2 + g^2A_3 + \dots   = e^{{\cal S}} \quad \rightarrow\quad  {\cal S} = \log {\cal A} = gS_1 + g^2S_2 + g^3S_3+ \dots
\end{equation}
where $S_L$ is a sum of $A_L$ and products of lower-loop $A$'s, for instance,
\begin{equation}
S_1 = A_1,\qquad S_2 = A_2 - \frac12A_1^2,\qquad S_3 = A_3 - A_2A_1 + \frac13 A_1^3, \quad etc. \label{logExp}
\end{equation}
The logarithm of the amplitude only has a mild log$^2$ divergence. This is reflected in a very special cut structure of its integrand $S_L$ as discussed in \cite{ArkaniHamed:2010gh} and more recently derived in \cite{Arkani-Hamed:2013kca} as following from amplituhedron geometry. 

The other object is the ratio function $R_{n,k}$. This is an IR-safe quantity defined as the ratio of N$^k$MHV amplitude to MHV amplitude. The expansion in loop order is then
\begin{equation}
{\cal R}^{(k)} = \frac{{\cal A}^{(k)}}{{\cal A}^{(0)}} = 1 + g R^{(k)}_1 + g^2 R^{(k)}_2 + \dots
\end{equation}
where
\begin{equation}
R_1^{(k)} = A_1^{(k)} - A^{(k)}_0A^{(0)}_1,\quad
R_2^{(k)} = A_2^{(k)} - A^{(k)}_0A^{(0)}_2 - A^{(k)}_1A^{(0)}_1 + A_0^{(k)}\left(A_1^{(0)}\right)^2,\quad etc.
\end{equation}

Let us look at the logarithm of the amplitude ${\cal S}$ in detail. The first non-trivial term in the expansion is $S_2$ where the result can be expressed using the same $Q_{ijkl}$ as in (\ref{DoublePent}), but where the ranges for indices are different:
\begin{equation}
S_2 = -\frac14 \sum_{i<j<k<l} Q_{ikjl} \label{LogL2}
\end{equation}
as was found in \cite{ArkaniHamed:2010gh} (the sum is here in a cyclic sense). This directly proves the positivity of $S_2$ in the positive region. All $Q_{ijkl}$ are now negative, the indices $j$ and $k$ are now in wrong order and the four-bracket $\la ijkl\ra$ changes the sign when put in canonical ordering. This minus sign is then compensated by the overall minus sign in (\ref{LogL2}) leaving the result manifestly positive. As an nice example we show $n=4$ case where the amplitude contains four double box integrals (which are just collapsed double pentagons $Q_{i\,i\pl1\,i\pl2\,i\pl3}$). The $S_2$ in that case can be written as
\begin{equation}
S_2 = \frac{\la 1234\ra^3(\la AB13\ra\la CD24\ra + \la AB24\ra\la CD13\ra)}{\la AB12\ra\la AB23\ra\la AB34\ra\la AB14\ra\la ABCD\ra\la CD12\ra\la CD23\ra\la CD34\ra\la CD14\ra} \label{Log4pt}\, ,
\end{equation}
where we used the Schouten identity
\begin{align}
&\hspace{-0.3cm}\la AB12\ra\la CD34\ra + \la AB23\ra\la CD14\ra + \la AB34\ra\la CD12\ra + \la AB14\ra\la CD23\ra - \la ABCD\ra\la 1234\ra\nonumber\\
&\hspace{6cm}= \la AB13\ra\la CD24\ra + \la AB24\ra\la CD13\ra \, .
\end{align}

The first four terms correspond to four double boxes in the two loop amplitudes while the last term is the one-loop square piece. This can be also seen from the structure of $Q_{ijkl}$ when we reshuffle indices: $Q_{1324}$ gives directly (\ref{Log4pt}). This numerator is manifestly positive in the amplituhedron.

Starting with $S_3$ it is harder to make the positivity manifest, since the full mutual positivity between the three loops comes into play. The expression for $S_3$ can be written as

\begin{equation}
S_3 =
\frac{\la1234\ra^3\left[N_3^{(a)} + N_3^{(b)} - N_{2,1}+ 2N_{1,1,1}\right]}{\begin{array}{c}\la AB12\ra\la AB23\ra\la AB34\ra\la AB14\ra\la ABCD\ra\la CD12\ra\la CD23\ra\la CD34\ra\\
\la CD14\ra\la EF12\ra\la EF23\ra\la EF34\ra\la EF14\ra\la ABEF\ra\la CDEF\ra\end{array}}\label{LogL3}
\end{equation}
where we denote by $AB$, $CD$, $EF$ three lines in $\mathbb{P}^3$ representing loop momenta, and we define building blocks coming from the expansion (\ref{logExp}):

\begin{align}
N_3^{(a)} & = \la 1234\ra\la ABEF\ra\la AB12\ra\la EF34\ra\la CD12\ra\la CD34\ra + S_{+1} + \sigma_{AB,CD,EF}\\
N_3^{(b)} &= \la AB12\ra^2\la CD23\ra\la CD34\ra\la EF34\ra\la EF14\ra + S_{+1,+2,+3} + \sigma_{AB,CD,EF}\\
N_{2,1} &= \la1234\ra\la AB12\ra\la CD34\ra\la ABEF\ra\la CDEF\ra + S_{+1} + \sigma_{AB,CD,EF}\\
N_{1,1,1}&=\la1234\ra^3 \la ABCD\ra\la ABEF\ra\la CDEF\ra
\end{align}

Here $S_{+1}$ stands for adding a cyclic term $Z_i\rightarrow Z_{i\pl1}$ and $Z_4\rightarrow Z_1$ while $S_{+1,+2,+3}$ is adding all three other cyclic terms. The symbol $\sigma_{AB,CD,EF}$ stands for summing over all permutations of lines $AB$, $CD$, $EF$. It is easy to see that all $N$'s are positive individually in the positive region. The minus sign generated by the cyclic shift in $\la1234\ra\rightarrow \la 2341\ra = - \la1234\ra$ is always compensated by the minus sign generated by $\la \ast\ast34\ra \rightarrow \la \ast\ast41\ra = - \la \ast\ast 14\ra$ once we write everything in the canonical ordering. It is very reasonable that there must be a minus sign in the numerator of (\ref{LogL3}) because all terms individually are ``too positive", in a sense that they are positive in bigger region than just the positive space for $L=3$ because the overlap of regions where $AB$, $CD$, $EF$ are mutually positive is not taken account. It would be still nice to see if there exists any way how to rewrite (\ref{LogL3}) as a sum of manifestly positive terms, perhaps with some ``less positive" building blocks \`{a} la \cite{Lam:2014jda} but it is also possible that the logarithm of the amplitude itself (\ref{LogL3}) is the smallest positive building block.

In the end we reverted to numerical checks and checked this expression to be positive as well as $S_3$ up to $n=8$ for $10^3$ kinematical points each.  For the case $n=4$ we checked the conjecture up to $L=5$ for $10^4$ points each and indeed the $S_L$ always stays positive in the positive region. For the integrand of the ratio function ${\cal R}^{(k)}$ we fully rely on numerical checks because the manifestly positive expansion does not exist for the same reason as for the integrand of the amplitude ${\cal A}^{(k)}$. We performed exactly the same numerical checks as for the amplitude with complete agreement with our conjecture.

It is already surprising that the integrand for the amplitude should be positive. Why would we expect the log of the MHV amplitude, or the general ratio function, to be positive? After all, these subtract from the integrand. But this is another qualitative feature that would follow from the existence of a ``dual" amplituhedron. Let's explain the intuition behind this in the simplest case of the polygon where everything is transparent. Starting with a polygon $Z_1,\cdots,Z_n$, let's add some $Z_{n+1}$. The polygon itself gets bigger: the new polygon with the point $Z_{n+1}$ added trivially contains the old one. But consider $\Omega$, for some $Y$ which is contained inside the first polygon and thus trivially inside the second one. It is obvious that while the polygon gets larger after adding the point, the form becomes {\it smaller}. This is clear even from the BCFW picture, where $Y$ is outside the extra triangle $(Z_n,Z_{n+1},Z_1)$, and thus the extra term in the triangulation is negative. It is even more obvious in terms of the area of the dual polytope, which gets {\it smaller} by chopping off a corner.
$$
\includegraphics[scale=0.65]{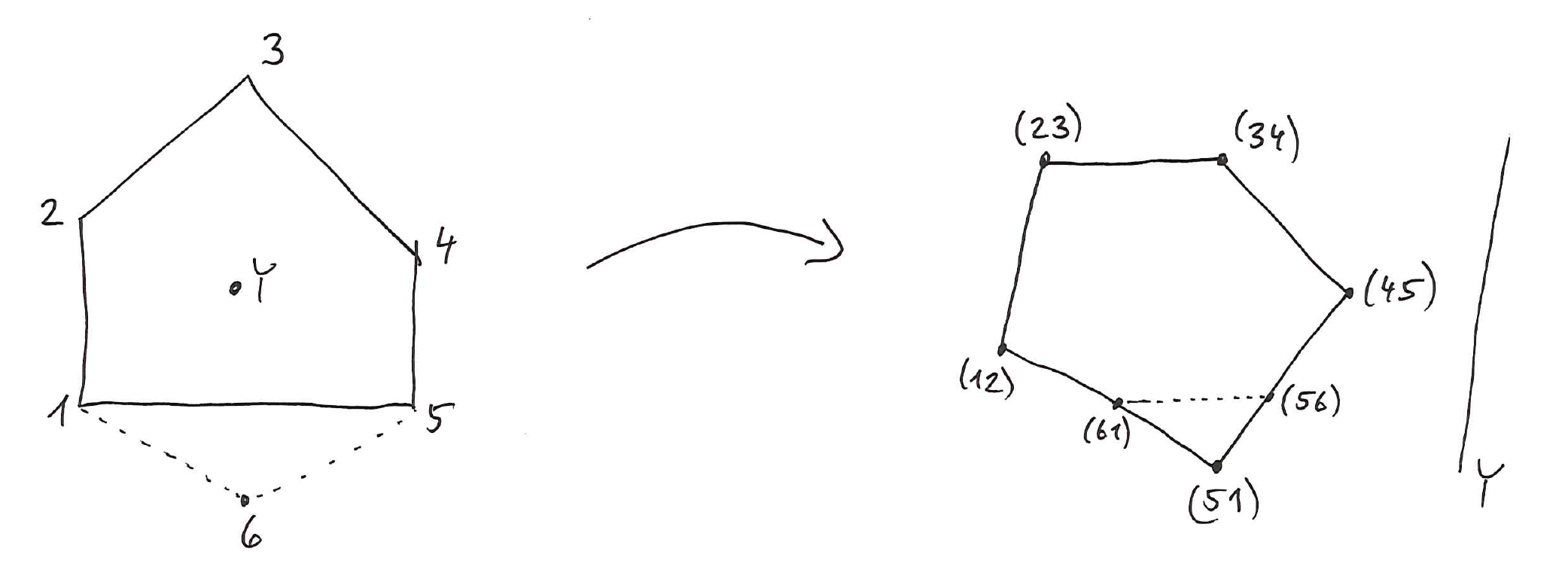}
$$
For the general amplituhedron, it is still obviously the case that the amplituhedron get larger when an extra point $Z_{n+1}$ is added. But we have made extensive numerical checks that, just as with the $k=1$ cases, the form $\Omega$ ${\it decreases}$, just as we would be qualitatively expect from a would-be-``dual amplituhedron" picture. More generally, this leads us to expect that a {\it larger} region in $Y$ space is associated with a {\it smaller} value for the form. This qualitative picture suggests that the log of the amplitude and the ratio function might have nice positivity properties.  Consider e.g.\ the log of the 2-loop amplitude. The region corresponding to ``1-loop $\times$ 1-loop" is ``larger", since we are imposing fewer positivity constraints than with the two-loop integrand. Since the region is larger, the corresponding form should be smaller, and thus subtracting it from the two-loop integrand to get the logarithm will leave us with something positive.

The positivity of the integrand for the ratio function also suggests that some interesting positivity might arise in the final amplitude, obtained after integration. (Note that positivity of the integrand inside the amplituhedron does not trivially imply this, since the standard contour of integration for the loop variables is not inside the amplituhedron in (2,2) signature, but over all of Minkowski space in (3,1) signature).

Of course after integration we no longer have the ``$AB$" variables, but the superamplitude can still be ``bosonized" in the $Y$-space of the amplituhedron. The expressions are given in terms of transcendental functions of cross-ratios weighted with Yangian invariants, and these can be easily uplifted to $Y$-space. We looked closely at the simplest example $k=1$ and $L=1$, for which the final result for the ratio function can be written as
\begin{equation}
Q_6 = H_1 \cdot \left[(2)-(3)+(4)\right]  + H_2 \cdot \left[(3)-(4)+(5)\right] + H_3\cdot \left[(4)-(5)+(6)\right]
\end{equation}
where $(1)=R[12345]$ is the $R$-invariant also familiar from the $k=1$ tree-level case,
\begin{equation}
R[abcde] = \frac{\la Yd^4Y\ra\la abcde\ra^4}{\la Yabcd\ra\la Ybcde\ra\la Ycdea\ra\la Ydeab\ra\la Yeabc\ra}
\end{equation}
and $H_1$ is the hexagon function
\begin{equation}
H_1 = \frac12\left[{\rm Li}_2 (1-u_1) + {\rm Li}_2 (1-u_2) + {\rm
Li}_2 (1-u_3) + \log(u_3)\log(u_1) - 2\zeta_2\right]\, .
\end{equation}

The cross-ratios are trivially upgraded to $Y$ space by adding a $Y$ to all the four-brackets:
\begin{equation}
u_1 = \frac{\la Y1234\ra\la Y4561\ra}{\la Y1245\ra\la Y3461\ra},\quad u_2 = \frac{\la Y2345\ra\la Y5612\ra}{\la Y2356\ra\la Y4512\ra},\quad
u_3 = \frac{\la Y3456\ra\la Y6123\ra}{\la Y3461\ra\la Y5623\ra}\, .
\end{equation}
All other cases $(j)$ and $H_i$ are related by cyclic shifts. In the end $Q_6$ is a form in $Y$ similar to the tree-level amplitude, but transcendental with rational pre-factors, rather than just rational. 

We can now take $Y$ inside the (tree) amplituhedron and test whether $Q_6$ is positive. An exhaustive check shows that indeed it is. This is quite non-trivial to show analytically: it does not simply follow from dilog identities but makes crucial use of the rational prefactors. These issues will be explored at greater length in  \cite{RatioPositivity}.

In addition to this case we also checked numerically the positivity of $R_1^{(1)}$ up to $n=10$ for $10^3$ points each using the Mathematica package \cite{Bourjaily:2013mma}, and found complete consistency with the positivity conjecture.  For $n=6$ the $L=2$ and $L=3$ cases will be discussed in \cite{RatioPositivity}, based on results obtained in \cite{Dixon:2011nj,Dixon:2014iba}.

It is also natural to investigate the positivity of the remainder function; here the (already indirect) connection with the positivity of the integrand is lost given that we don't have an integral representation of the log of the amplitude upon subtracting the BDS term; nonetheless it is certainly interesting to explore the positivity properties of the remainder function for positive external data as well. This will be investigated at multiloop order in \cite{RatioPositivity}.

\section*{Acknowledgements}

We thank Jake Bourjaily and Thomas Lam for stimulating discussions. We also thank Simon Caron-Huot for discussions, and for suggesting and checking the positivity of the integrated 6-point ratio function at one loop.  N.~A.-H. is supported by the Department of Energy under grant number DE-FG02-91ER40654. J.~T. is supported in part by the David and Ellen Lee Postdoctoral Scholarship and by the Department of Energy under grant number de-sc0011632.

\end{document}